\documentclass[aps,pre,reprint,showpacs,superscriptaddress]{revtex4}
\usepackage{graphicx,color,epsfig}
\usepackage{amssymb,amsmath,amsfonts}

\begin{document}

\title{Synchronization and phase ordering in globally coupled chaotic maps}
\author{O. Alvarez-Llamoza}
\affiliation{Departamento de F\'isica, FACYT, Universidad de Carabobo, Valencia, Venezuela}
\author{M. G. Cosenza}
\affiliation{Centro de F\'isica Fundamental, Universidad de Los Andes, M\'erida, Venezuela}
%\date{ }

\begin{abstract}
We investigate the processes of synchronization and phase ordering 
in a system of globally coupled maps possessing bistable, chaotic local dynamics. 
The stability boundaries of the  synchronized 
states are determined on the space of parameters of the system.  
The collective properties of the system are characterized by means of the persistence probability of equivalent spin variables
that define two phases, and by a magnetization-like order parameter that measures the phase-ordering behavior.
As a consequence of the global interaction, the persistence probability saturates 
for all values of the coupling parameter, in contrast to the 
transition observed in the temporal behavior of the persistence in
coupled maps on regular lattices. A discontinuous transition from a non-ordered state to a collective phase-ordered state 
takes place at a critical value of the coupling. On an interval of the coupling parameter, we find three distinct realizations 
of the phase-ordered state, which can be discerned by the corresponding values of the saturation persistence. Thus, 
this statistical quantity can provide information about the transient behaviors that lead to the different phase configurations in the system. 
The appearance of disordered and phase-ordered states in the globally coupled system can be understood by calculating 
histograms and the time evolution of local map variables associated to the these collective states.
\end{abstract}


\maketitle

\section{Introduction.}
Globally coupled systems have been a research topic receiving a large amount of attention because of their 
applicability to a variety of contexts. 
The dynamical elements in such systems are subject to a common interaction field. 
Global interactions arise in the description of many physical,
biological, chemical and social systems, such as Josephson junction arrays \cite{Hadley}, multimode lasers \cite{Wie}, coupled oscillators \cite{Kuramoto,Nakagawa}, charge density waves \cite{Gruner},  parallel electric circuits, neural
dynamics, ecological systems, evolution models \cite{Kaneko1}, economic
exchange \cite{Yakovenko}, social networks \cite{Newman}, mass media models \cite{Media},
cross-cultural interactions \cite{Plos}, etc. 
Global interactions also play a relevant role in
models of many systems driven by long-range interactions, able to generate strong correlations between highly
interconnected elements. Systems with global interactions can exhibit a variety of phenomena, 
such as chaos synchronization,  nontrivial
collective behavior, dynamical clustering, chaotic itineracy \cite{Kaneko1,Manrubia}, quorum sensing \cite{Ojalvo}, etc. 
These behaviors have been experimentally investigated in arrays of globally coupled oscillators in several systems 
\cite{Wang,Yamada, DeMonte,Taylor,Roy}.

In addition to these phenomena,  
the description of generic effects associated to the presence 
of global coupling in dynamical processes in complex systems               
is still an open problem. In this respect, globally coupled maps \cite{Kaneko2} constitute paradigmatic models for the study of 
dynamical systems with global interactions. 
Spatial concepts lose meaning and only temporal properties
become relevant in globally coupled maps. These characteristics should introduce new features
in many processes that have been investigated in spatially extended systems with short range interactions. 

In particular, there has been much interest in the study of the phase-ordering properties of
systems of coupled bistable maps and their relationship with
Ising models in statistical physics \cite{Huse,Hern,Chate,Kock,Liu,Just,Pelli,Angel,Tucci,Ceche}. These works have mainly assumed the phase
competition dynamics taking place on networks with local interactions. 

In this article we investigate the collective behavior of a system of globally coupled bistable chaotic maps,
including the occurrence of synchronized states and 
the phenomenon of phase competition. 
This model provides a scenario to compare the roles that local and global
interactions play on the occurrence of phase growth and phase transitions on spatiotemporal systems. 
In Sec.~(\ref{II}) we present the model of globally coupled maps and describe local dynamics that exhibits
bistable, chaotic behavior. In Sec.~(\ref{III}) we determine analytically the stability condition for synchronized states on 
the space of parameters of the system. The phase-ordering properties associated to the collective dynamics of the system
are studied in Sec.~(\ref{IV}) by employing appropriate statistical quantities. Section~(\ref{V}) contains the conclusions of this work.  

\section{Globally coupled bistable chaotic maps.}
\label{II}
We consider a globally coupled map system defined by
\begin{equation}
 x^i_{t+1}=(1-\varepsilon)f(x^i_t)+\frac{\varepsilon}{N} \sum^N_{j=1} f(x^j_t),
\label{global}
\end{equation}
where $x^i_t$ describes the state variable of element $i$ ($i=1,2,\ldots,N$), at discrete time $t$, 
the parameter $\varepsilon$ measures the coupling strength between the elements,
and $f(x)$ is a map that expresses the local dynamics. The term expressing the global coupling between the maps 
corresponds to the mean field of the system. In this article, we employ a system size $N=10^5$.

The local dynamics is given by a piecewise linear, odd map 
\begin{equation}
  f(x)=\left\{
    \begin{tabular}{ll}
      $-2\mu /3-\mu x$, & if $x$ $\in $ $\left[ -1,-1/3\right] , $ \\
      $\mu x$, & if $x$ $\in $ $\left[ -1/3,1/3\right], $ \\
      $2\mu /3-\mu x$, & if $x$ $\in $ $\left[ 1/3,1\right]$ ,
    \end{tabular}
  \right. \label{sierra}
\end{equation}
where the local parameter $\mu \in [-3,3]$ and  $x \in [-1,1]$. 
For $\mu=3$, $f(x)$ becomes the chaotic map introduced by
Miller and Huse \cite{Huse}. For $\mu \in [-1,1]$, the map possesses the stable fixed point $f(x^*)=x^*=0$.
When the parameter $\mu \in (1,2)$, the local map is chaotic and bistable: there are two symmetric chaotic band attractors, 
corresponding to the invariant intervals $I^\pm=[\pm\mu(2-\mu)/3, \pm \mu/3]$, and separated by a finite gap about the origin. 
This map has been shown to exhibit phase-ordering properties
on locally coupled map lattices \cite{Chate,Kock}.

\section{Synchronized states.}
\label{III}
The coupled map system Ec.~(\ref{global}) can be expressed in vector form as 
\begin{equation}
\label{matriz}
{\mathbf x}_{t+1}=(1-\varepsilon) {\mathbf f}(x_t)+ \frac \varepsilon N  {\mathbf {M f}}(x_t)=\left[ (1-\varepsilon) 
{\mathbf I}+ \frac \varepsilon N  {\mathbf M} \right] {\mathbf f}(x_t),
\end{equation}
where ${\mathbf x}_t$ and ${\mathbf f}(x_t)$ are $N$-dimensional vectors with components 
$[{\mathbf x}_t]_i=x^i_t$ and $[{\mathbf f}(x_t)]_i=f(x^i_t)$, respectively; ${\mathbf I}$ is the  $N \times N$ identity matrix; 
and ${\mathbf M}$ is an $N \times N$ matrix expressing the coupling between the elements. For the global coupling, Ec.~(\ref{global}), 
all the components of ${\mathbf M}$ are equal to $1$.

A synchronized state occurs when $x^i_t=x^j_t$, $\forall \, i, j$. From the linear stability analysis of synchronized states 
in coupled map lattices, it can be shown that these states are stable if the following condition is satisfied \cite{Kaneko2,waller},
\begin{equation}
  \label{sinc}
  \left| \left(1-\varepsilon + \frac \varepsilon N m_k \right) e^{\lambda} \right| <1,
\end{equation}
where $\{ m_k: k=1, 2, \ldots, N\}$ is the set of eigenvalues of the coupling matrix ${\mathbf M}$ and $\lambda$ 
is the Lyapunov exponent of
the local map, Ec.~(\ref{sierra}). In the globally coupled case, the eigenvalues are
$m_k=0$, $k=1, \ldots,(N-1)$,  which has $(N-1)$-fold degeneracy, and $m_N=N$. 
Because of these eigenvalues, the synchronization condition, Ec.~(\ref{sinc}), is independent of the size of the system $N$. 

The  set of  eigenvectors of the matrix ${\mathbf M}$ constitute a complete orthogonal basis in terms of 
which any state ${\mathbf x}_t$ of the system Ec.~(\ref{matriz}) can be represented as a linear combination. 
The eigenvector corresponding to $m_N=N$ is homogeneous 
and it expresses the coherent or synchronized state at any time.
Thus, perturbations of the state ${\mathbf x}_t$ along this eigenvector do not destroy the coherence, and the stability 
condition associated with $m_N=N$ is irrelevant for the synchronized state. The other $(N-1)$ eigenvectors associated with 
$m_k=0$ are not homogeneous, 
and perturbations along their directions affect the synchronized state. 
Thus, the stability condition Ec.~(\ref{sinc}) with $m_k=0$ defines a region on the space of parameters $(\mu,\varepsilon)$ 
where all the stable synchronized states can be found.

Two types of synchronized states fulfilling condition Ec.~(\ref{sinc})  can be observed in the system:
\begin{enumerate}
\item Synchronized stationary states,  for which $x^i_t=x^*, ~ \forall ~ i$. 
This corresponds to the range of parameter $\mu \in [-1,1]$, where the local map possesses the
stable fixed point $f(x^*)=x^*=0$. The boundaries of the region of parameters where this state is stable are given by
Ec.~(\ref{sinc}) with $\mu_k=0$ and $e^\lambda= f'(x^*)$, 
\begin{equation}
  \label{sinces}
   \left(1-\varepsilon \right)  |f'(x^*)| =\pm 1.
\end{equation}
\item Synchronized chaotic states, for which $x^i_t=f(x_t), ~ \forall ~ i$. This occurs in the 
regions $\mu \in [-3,-1] ~\cup ~  [1,3]$,
where the local map is chaotic. The region of stability of these states is bounded by the curves
\begin{equation}
  \label{sincca}
  \left(1-\varepsilon  \right) e^{\lambda} =\pm 1,
\end{equation}
\end{enumerate}
Since $f'(x^*)=\mu$ and $\lambda= \ln |\mu|$ for the map Ec.~(\ref{sierra}), both boundaries (\ref{sinces}) and (\ref{sincca}) 
can be expressed on the space of parameters $(\mu,\varepsilon)$ by the curves
\begin{equation}
\label{bound}
  \left(1-\varepsilon  \right) |\mu |=\pm 1, 
\end{equation}
with $\mu$ in the appropriate range for each state. The straight lines $\mu=-1$ and $\mu=1$ separate the synchronized stationary states from
the synchronized chaotic states on the plane $(\mu,\varepsilon)$.

The occurrence of synchronization can also be numerically characterized by the asymptotic time-average
$\langle\sigma\rangle$ (after discarding a number of transients) of the instantaneous standard deviations
$\sigma_t$ of the distribution of state variables $x^i_t$, defined as
\begin{equation}
\label{sigma}
\sigma_t=\left[ \frac{1}{N} \sum_{i=1}^N \left( x^i_t - \bar x_t \right)^2 \right]^{1/2} ,
\end{equation}
where  
\begin{equation}
\label{mean}
\bar x_t=\frac{1}{N} \sum^N_{i=1} x^i_t \, .
\end{equation}
Then, a synchronization state corresponds to a value $\langle \sigma \rangle=0$. In practice, we use the numerical criterion
$\langle \sigma \rangle < 10^{−7}$ as a synchronization condition.

For some values of parameters, the iterates of the state variables $x^i_t$ in the system Ec.~(\ref{global}) leave the interval $[-1,1]$ and,
eventually, escape to infinity. The iterates of $x^i_t$ stay in the interval $[-1,1]$ if the product $(1-\varepsilon)\mu$ lies in the range $[-3,3]$;
that is if $|(1-\varepsilon)\mu| < 3$. Thus, the boundaries for escape from the interval $[-1,1]$ are described by the curves
\begin{equation}
\label{escape}
  \left(1-\varepsilon  \right) \mu =\pm 3.
\end{equation}

\begin{figure}[h]
\centerline{
\includegraphics[scale=0.32,angle=0,trim=2 2 2 2,clip]{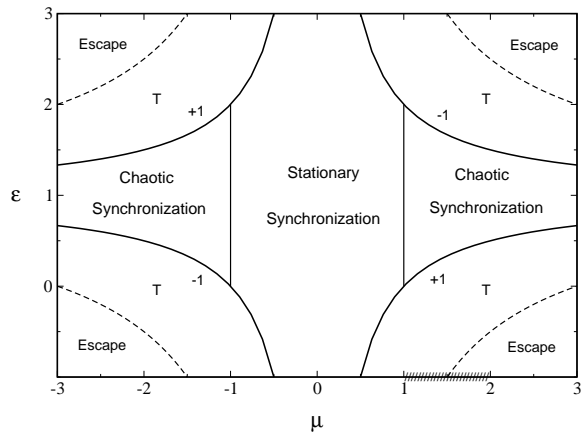}}
\caption{Regions of stable synchronized states for the system Ec.~(\ref{global}) on the space of parameters $(\mu, \varepsilon)$. 
The labels indicate where the synchronized stationary states and the synchronized chaotic states occur.
The synchronization boundaries correspond to continuous lines. The labels $\pm 1$
on each curve identify the corresponding sign in Ec.~(\ref{bound}). The dashed lines indicate the escape boundaries Ec.~(\ref{escape}) beyond which
the iterates of the state variables of the system Ec.~(\ref{global}) leave the interval $[-1,1]$.
The interval $\mu \in [1,2]$ for bistability is marked on the $\mu$ axis. The label {\sf T} identifies the regions  where  collective turbulent states exist.} 
\label{frontiers}
\end{figure}

Figure~(\ref{frontiers}) shows the stability boundaries of the synchronized states and the escape boundaries for the globally coupled system
Ec.~(\ref{global}) on the space of parameters $(\mu, \varepsilon)$. 

\section{Collective phases.}
\label{IV}
For $\mu \in (1,2)$,
the local map displays  bistability in the form
of two chaotic band attractors: corresponding to 
the interval $I^+$ for the positive values of the iterates, and the interval $I^-$ for the negatives values. 
Then the states of the elements in the system Ec.~(\ref{global}) can be associated to two well defined 
symmetric phases that can be characterized by spin variables
associated to the sign of the state at time $t$,
defined as $s^i_t = +1$ if $x^i_t >0$, and  $s^i_t = -1$ if $x^i_t <0$.

To study the collective behavior of the globally coupled
map system Ec.~(\ref{global}) in the  bistable chaotic range,
we fix the value of the local parameter $\mu=1.9$ and choose an even number $N$ as system size. 
Then we set the initial conditions symmetrically as follows: one half of
the maps are randomly chosen and assigned random values uniformly distributed on the
positive attractor while the other half are similarly assigned values on the negative attractor.

\begin{figure}[h]
\centerline{
\includegraphics[scale=0.32,angle=0,trim=2 2 2 2,clip]{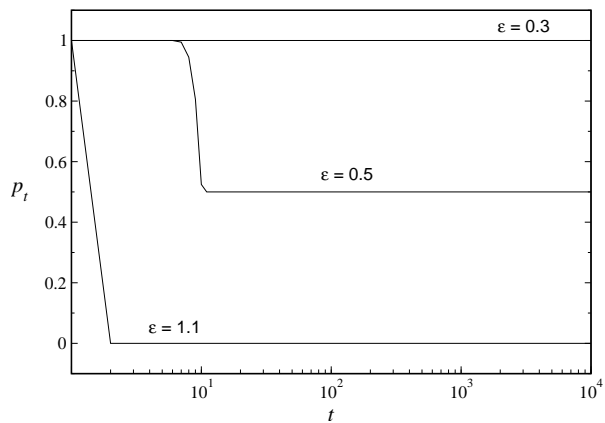}}
\caption{Persistence probability as a function of time for the system Ec.~(\ref{global}) with fixed $\mu=1.9$ and size $N=10^5$, 
for different values of the coupling parameter $\varepsilon$, as indicated on each curve.}
\label{pertu1.9}
\end{figure}

The dynamical properties of the phase-ordering process can be described by using the persistence probability $p_t$,
defined as the fraction of maps that have not changed spin variable (sign) up to time $t$ \cite{Derrida}. 
Figure~(\ref{pertu1.9}) shows $p_t$ as a function of time for the globally coupled
map system Ec.~(\ref{global}), for several values of the coupling parameter $\varepsilon$.  
The persistence probability saturates in a few iterations for all positive values of the coupling $\varepsilon$. 
This means that the phases associated to the spin variables freeze in the globally coupled system.
In contrast, in regular lattices the persistence saturates for small couplings, while it decays algebraically in time 
for coupling strengths greater than some critical value, corresponding to the growth of one phase at the expense of the other \cite{Chate}. \\

\begin{figure}[h]
\centerline{
\includegraphics[scale=0.7,angle=0]{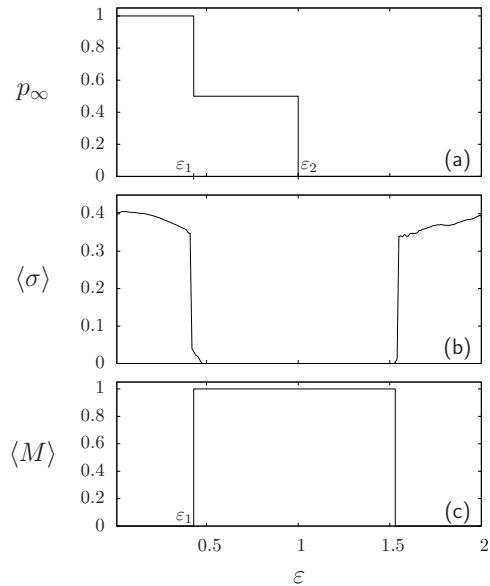}}
\caption{Statistical quantities 
as functions of the coupling parameter $\varepsilon$ for  system Ec.~(\ref{global}) with fixed $\mu=1.9$ and size $N=10^5$. 
(a) $p_\infty$; (b) $\langle \sigma \rangle$; (c) $\langle M \rangle$. 
The critical values $\varepsilon_1$ and $\varepsilon_2$ are marked on the $\varepsilon$ axis.}
\label{pertasim}
\end{figure}

Figure~(\ref{pertu1.9}) reveals that the saturation value of the persistence probability, denoted by $p_\infty$, 
depends on the value of the coupling parameter. 
Figure~(\ref{pertasim}a) shows the quantity $p_\infty$ as a function of $\varepsilon$.  
We find that $p_\infty$ displays different constant values in different intervals of the coupling parameter and exhibits discontinuous transitions
at critical values  $\varepsilon_1=0.43$ and $\varepsilon_2=1$. 
For $\varepsilon < \varepsilon_1$, we have $p_\infty=1$, indicating that for small enough coupling, every map remains in its initial chaotic attractor,
$I^-$ or $I^+$. In the intermediate range of coupling parameters $\varepsilon_1 < \varepsilon < 1$, the saturation value of the persistence
changes to $p_\infty=0.5$, indicating that one-half of the total number of maps have switched attractor. Finally, for $\varepsilon > 1$, we obtain
$p_\infty=0$; this means that all the maps have changed their initial attractors at some time during the evolution of the system. 
The value of $\varepsilon_1$ depends on the value of the local map parameter $\mu$, but  $\varepsilon_2=1$, independently of $\mu$.

Figure~(\ref{pertasim}b) shows the synchronization measure $\langle \sigma \rangle$, given by Eq.~(\ref{sigma}), as a function of
$\varepsilon$. 
For the value $\mu=1.9$, the chaotic synchronization range of the coupling, obtained from Eq.~(\ref{bound}), is
$\varepsilon \in [0.473,1.526]$. For this interval of the coupling, we get $\langle \sigma \rangle=0$ as expected.

To characterize the statistical properties of the phase-ordering process in the globally coupled
map system Ec.~(\ref{global}), 
we define the instantaneous ``magnetization'' of the system $M_t$, as 
\begin{equation}
M_t = \frac{1}{N} \sum_{i=1}^N s_t^i. 
\end{equation}
Then, we employ, as an order parameter, the absolute value of the asymptotic time-average 
(after discarding a number of transients) of the values $M_t$, denoted by $\langle M \rangle$. 

Figure~(\ref{pertasim}c) shows the order parameter $\langle M \rangle$ as a function of the coupling  $\varepsilon$.
The critical value of the coupling $\varepsilon_1$ marks a discontinuous transition from a collective state characterized by $\langle M \rangle=0$, where the maps remain symmetrically distributed about the two attracting intervals $I^+$ and $I^-$, to an ordered state characterized by 
$\langle M \rangle=1$, where all the maps settle on one of the attractors, either $I^+$ or $I^-$. 
Note that the critical value $\varepsilon_1$ is smaller than the lower synchronization boundary at $\varepsilon=0.473$. 
This means that the phase-ordering transition occurs before full synchronization is achieved. 
When the value of the coupling strength reaches
the upper synchronization boundary, there is another discontinuous transition 
to the turbulent, disordered state, for which $\langle M \rangle=0$.

The quantities $p_\infty$ and $\langle \sigma \rangle$ in Fig.~(\ref{pertasim}) allow to distinguish 
three different situations in the range of the parameter $\varepsilon$ where the collective phase-ordered state with $\langle M \rangle=1$ occurs:
(i) a desynchronized ordered state, where $\langle M \rangle=1$, $\langle \sigma \rangle >0$, and  $p_\infty=0.5$; 
(ii)  a synchronized ordered state, where $\langle M \rangle=1$, $\langle \sigma \rangle=0$, and $p_\infty=0.5$; and
(iii) a synchronized ordered state, characterized by $\langle M \rangle=1$, $\langle \sigma \rangle=0$, and $p_\infty=0$.

\begin{figure}[h]
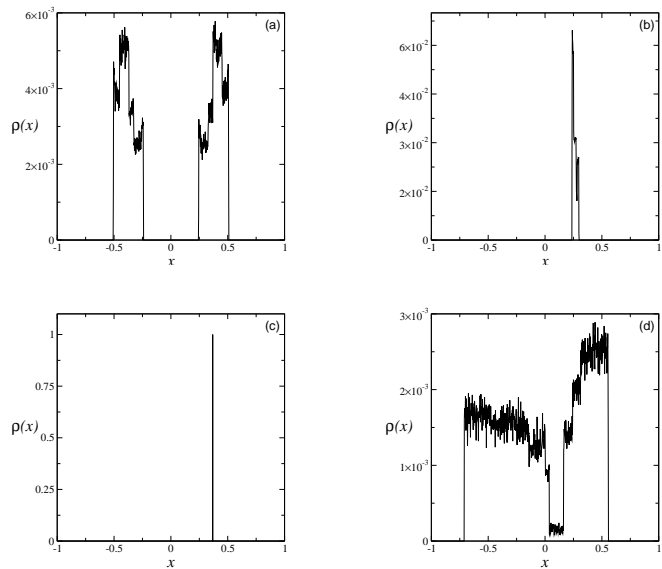

\centerline{
\includegraphics[scale=0.2,angle=0]{Figure4a.eps}
\hspace{0.8cm}
\includegraphics[scale=0.2,angle=0]{Figure4b.eps}}
\vspace{0.5cm}
\centerline{
\includegraphics[scale=0.2,angle=0]{Figure4c.eps}
\hspace{0.8cm}
\includegraphics[scale=0.2,angle=0]{Figure4d.eps}}
\caption{Instantaneous probability distributions $\rho(x)$ of states at time $t=2000$ for the system Ec.~(\ref{global}) with  $N=10^5$ 
and fixed $\mu=1.9$, for different values of the coupling $\varepsilon$. 
(a) $\varepsilon=0.2$ (non-ordered phase, $p_\infty=1$, $\langle M \rangle=0$), (b) $\varepsilon=0.45$ (desynchronized ordered phase, $\langle \sigma \rangle >0$, $p_\infty=0.5$, $\langle M \rangle=1$), (c) $\varepsilon=0.55$ (synchronized ordered phase,  
$\langle \sigma \rangle=0$, $p_\infty=0.5$, $\langle M \rangle=1$). 
(d) $\varepsilon=2.4$ (turbulent).}
\label{histog}
\end{figure}

In order to elucidate the nature of these three realizations of the phase-ordered state, as well as
the transitions exhibited by the statistical quantities $p_\infty$ and $\langle M \rangle$,  we plot in
Fig.~(\ref{histog}) the instantaneous probability distributions 
of the states $x^i_t$ of the system Eq.~(\ref{global}) with fixed $\mu=1.9$, denoted by $\rho(x)$, 
for different values of the coupling parameter $\varepsilon$.   

\begin{figure}[h]
\centerline{
\includegraphics[scale=0.58,angle=0]{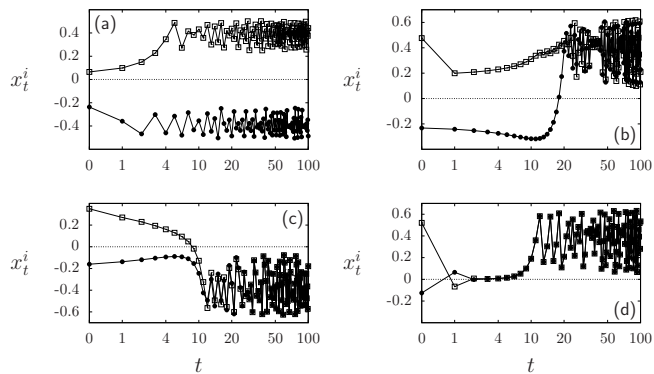}}
\caption{Temporal evolution of the state variables $x_t^i$ of a map with positive initial spin variable (empty squares) and a map with negative
initial spin variable (black circles) in the globally coupled system Ec.~(\ref{global}) with  $N=10^5$ and fixed $\mu=1.9$, for different values
of the coupling $\varepsilon$. For $t>0$, time is displayed in a logarithmic scale. 
(a) $\varepsilon=0.2$; non-ordered state, with $\langle \sigma \rangle>0$, $p_\infty=1$, $\langle M \rangle=0$.
(b) $\varepsilon=0.45$; desynchronized, ordered state, with  $\langle \sigma \rangle>0$, $p_\infty=0.5$, $\langle M \rangle=1$.
(c) $\varepsilon=0.55$; synchronized, ordered state, with $\langle \sigma \rangle=0$, $p_\infty=0.5$, $\langle M \rangle=1$.
(d) $\varepsilon=1.1$; synchronized, ordered state, with  $\langle \sigma \rangle=0$, $p_\infty=0$, $\langle M \rangle=1$. }
\label{frontiers2}
\end{figure}

Figure~(\ref{histog}a) corresponds to  $\varepsilon=0.2 < \varepsilon_1$. 
The  probability distribution $\rho(x)$ at $t=2000$ shows two separated peaks that maintain the initial symmetrical distribution of the maps on
the two chaotic band attractors. The mean field coupling term is negligible in this situation
Figure~(\ref{frontiers2}a) shows the time evolution of the state variables $x_t^i$ of two maps in the system Ec.~(\ref{global}) for  $\varepsilon=0.2$: one having positive initial spin variable and another with negative initial spin variable. Each trajectory remains in its attractor. Since no map has left its initial attractor, 
no spin variable has changed sign, and thus $p_\infty=1$. Then, two symmetric subsets associated to the spin variables coexist in the globally coupled system for these parameters values, yielding
$\langle M \rangle=0$ as a result. 

For couplings $\varepsilon_1 < \varepsilon < 0.473$ in Fig.~(\ref{histog}b), the probability distribution $\rho(x)$ at time $t=2000$ 
displays one single peak. This indicates that the $N/2$ maps that initially belonged to one attractor have switched to the other attractor; 
the direction of the change depends on the initial conditions, in this case
from $I^-$ to $I^+$. All the maps form a cluster that moves chaotically and stays in the interval $I^+$. 
Then, the saturation value of the persistence probability is $p_\infty=0.5$ in this range of the coupling strength. 
Figure~(\ref{frontiers2}b) illustrates this process through the time evolution of the orbits of two maps that have been initially assigned
opposite spin variables. Note that the two chaotic orbits do not synchronize on the interval $I^+$. 
This situation corresponds to 
the presence of a single ordered phase of spin variables in the system, and therefore the magnetization becomes $\langle M \rangle=1$.
The discontinuous change in the statistical quantities 
occurring at the critical value $\varepsilon_1$ in Fig.~(\ref{pertasim}) describes a first order phase transition in the collective behavior of the system, from a non-ordered  state, characterized by the values $\langle \sigma \rangle >0$, $p_\infty=1$, $\langle M \rangle=0$, 
to a desynchronized  phase-ordered state, characterized by $\langle \sigma \rangle >0$, $p_\infty=0.5$, $\langle M \rangle=1$, and denoted as
situation (i) above.

For coupling values $0.473 < \varepsilon < 1$, 
the probability  $\rho(x)$ at  $t=2000$ displays a single vertical line on one of the attracting intervals, as shown in Fig.~(\ref{histog}c).
This indicates that the $N/2$ maps initially assigned to one of the attractors have switched to the other attractor,
resulting in the synchronization of the $N$ maps on a single chaotic orbit that stays on that attractor. 
The corresponding time evolution of two maps with initial opposite spin variables is shown in Fig.~(\ref{frontiers2}c).
Thus, the system  displays a synchronized, phase-ordered state characterized by $\langle \sigma \rangle=0$,
$p_\infty=0.5$, and $\langle M \rangle=1$ in this parameter range. This constitutes realization (ii) of the phase-ordered state.

If the coupling is increased to values $1 < \varepsilon < 1.526$, we observe realization (iii) of the ordered state.
In this case, the factor $(1-\varepsilon)$ in Ec.~(\ref{global}) becomes negative 
allowing the maps to reverse the signs of their initial spin variables 
at early times during the evolution of the system. 
This transient behavior of the spin variables is reflected in the quantity $p_\infty=0$; the parameter $ \varepsilon_2=1$ 
marks the discontinuity in the value of the saturation value of the persistence. 
Since the synchronized state is stable for this range of coupling parameters, the maps 
eventually become synchronized on one of the attracting intervals, yielding 
$\langle \sigma \rangle =0$. 
In addition, we obtain $\langle M \rangle=1$.  Figure~(\ref{frontiers2}d) portrays the time evolution of the orbits of two maps with different initial spin variables in this situation. 
Then, we  have again a synchronized, phase-ordered state, distinguished by the value
$p_\infty=0$, in contrast to realization (ii) where $p_\infty=0.5$. Thus, the saturation value of the persistence probability 
can provide information about the transient processes that lead to frozen configurations and to phase-ordered states in the system.

In Fig.~(\ref{histog}d) for $\varepsilon > 1.526$, the maps become desynchronized and the corresponding probability distribution 
$\rho(x)$ is spread over a 
a subset of the interval $x \in [1,1]$; its bimodal form reflects the underlying presence of the two attractors in the local chaotic dynamics.
We refer to this collective state as turbulent. This corresponds to a desynchronized, disordered state for which 
$\langle \sigma \rangle >0$, $p_\infty=0$, and $\langle M \rangle=0$.

If the initial conditions are modified in such a way that a fraction $N_1/N > 1/2$ of values of the maps is uniformly distributed on one attractor interval,
either $I^+$ or $I^-$, while
the remaining fraction $(1-N_1/N)$ is similarly assigned to the other attractor, then the symmetry of the globally coupled system
Ec.~(\ref{global}) is lost. We have found that the main features of the collective behavior of the system are maintained under such partition:
the persistence probability $p_t$ saturates after a few iterations for all values of the coupling parameter;  
there is a disordered state for low coupling values; and a synchronized phase-ordered state emerges for an intermediate range of the coupling strength. 

In this situation, the mean field of the system initially acquires the sign of the attractor where the largest fraction  
$N_1/N$ of maps  lies. This attractor dominates the dynamics of the globally coupled bistable system. However, for low enough intensity of the coupling, 
the maps tend to stay in their initial intervals and therefore they do not change the sign of their spin variable. 
Correspondingly, the
asymptotic persistence probability has the value  $p_\infty=1$ and the average magnetization is $\langle M \rangle =0$. 
As the coupling strength is increased, 
the maps in the smallest subset  switch to the dominating attractor,  eventually giving rise to a phase-ordered state.
The initial fraction of maps $N_1/N$ remain on that attractor and therefore do not change sign in their spin variables. Then, the  saturation value of the persistence 
probability for ordered state in this case should be $p_\infty=N_1/N$. We have numerically verified these values for different partitions 
$(N_1, N-N_1)$ of initial conditions over the attracting intervals.

\section{Conclusions.}
\label{V}
We have investigated the collective behavior of a system of globally coupled maps having bistable, chaotic local dynamics. 
The system possesses two types of synchronized dynamics: synchronized stationary states and synchronized chaotic states. 
We have analytically determined the stability boundaries of these states on the space of parameters $(\varepsilon, \mu)$ of the system.  
In addition, we have numerically measured the occurrence of synchronization by means of the statistical quantity $\langle \sigma \rangle$.

The presence of two symmetric attracting intervals in the local chaotic dynamics permits to assign a spin-like variable to each map and to define associated phases. The persistence probability $p_t$ describes the evolution and competition of the phases. 
The absence of spatial relations in the system of globally coupled 
maps rules out the possibility of supporting spatial domains in either phase and a defined interface which would be necessary for 
a continuous phase growth.  
As a consequence, the phases always freeze in globally coupled maps, causing the saturation of the persistence probability in time
for all values of the coupling parameter, in contrast to 
the transition observed in the temporal behavior of the persistence in
coupled maps on regular lattices. We have found that the  saturation value of the persistence probability $p_\infty$ 
reaches different constant values in different intervals of the coupling parameter and shows discontinuous transitions
at critical values  $\varepsilon_1=0.43$ and $\varepsilon_2=1$.

We have introduced the magnetization-like order parameter $\langle M \rangle$ to characterize the phase-ordering behavior of the system. 
The phase-ordered state, corresponding to $\langle M \rangle=1$, exhibits three distinct realizations as the coupling  $\varepsilon$ is
varied and which can be discerned by employing the quantities $\langle \sigma \rangle$ and  $p_\infty$: 
(i) a desynchronized ordered state, with $\langle M \rangle=1$, $\langle \sigma \rangle >0$, and  $p_\infty=0.5$; 
(ii) a synchronized ordered state, characterized by $\langle M \rangle=1$, $\langle \sigma \rangle=0$, and $p_\infty=0.5$; and
(iii) a synchronized ordered state, distinguished by $\langle M \rangle=1$, $\langle \sigma \rangle=0$, and $p_\infty=0$.
Thus, the value of $\langle \sigma \rangle$ distinguishes between realizations (i) and (ii); while $p_\infty$ differentiates
realization (ii) from realization (iii).

There also exist two desynchronized, non-ordered collective states, both described by the values 
$\langle \sigma \rangle>0$ and $\langle M \rangle=0$.
One of these states corresponds to the persistence of the initial symmetric distribution of spin variables, characterized by $p_\infty=1$; and the other is a turbulent state, where $p_\infty=0$. Our results reveal that the saturation value of the persistence probability can provide information  
about the transient behaviors that lead to the different phase configurations in the system.  

In addition, we have studied the histograms and the time evolution of local map variables associated to the
disordered and to the phase-ordered states in order to understand the appearance of these collective states in the globally coupled system. 
The transitions between the disordered states and the phase-order state occur
discontinuously, as in a first-order phase transition, reflecting the global nature of the interaction in the system.
In contrast, continuous transitions are typical in regular lattices. 

In general, the investigation of dynamical processes in networks and the role of the topology in determining emerging collective behaviors
is a topic of much interest in the current research on complex systems \cite{Newman,Herrera}. 

\section*{Acknowledgements} 
This work is supported by project No. C-1827-13-05-B from CDCHTA, 
Universidad de Los Andes, Venezuela. M. G. C. is grateful to the Senior Associates Program of 
the Abdus Salam International Centre for Theoretical Physics, Trieste, Italy.

\end{document}